\documentclass{article}
\usepackage{spconf,amsmath,graphicx,bm,multicol,multirow, url}
\usepackage{booktabs}
\usepackage{adjustbox}
\usepackage[table, svgnames, dvipsnames]{xcolor}
\usepackage{hhline}
\definecolor{salmon}{HTML}{FFCCC9}
\usepackage{amssymb}
\newcommand{\argmax}{\operatornamewithlimits{arg\,max}}
\usepackage{tikz}
\usepackage{enumitem}
\setlist[enumerate]{leftmargin=*}
\usepackage[T1]{fontenc}

\definecolor{our_green}{HTML}{74af38}
\definecolor{blue1}{HTML}{7777ff}

\definecolor{lv_green}{HTML}{008080}
\definecolor{lv_blue}{HTML}{00008B}

\definecolor{compare_green}{HTML}{A3B6A2}

\definecolor{compare_light_blue}{HTML}{94ACCA}
\definecolor{compare_dark_blue}{HTML}{6767CD}

\usepackage[labelformat=simple]{subcaption}

\usepackage[font=small,labelfont=bf]{caption}

\title{Learning from Flawed Data: Weakly Supervised \\Automatic Speech Recognition}
\name{Dongji Gao$^1$, Hainan Xu$^3$, Desh Raj$^1$, Leibny Paola Garcia Perera$^{1,2}$, 
Daniel Povey$^4$, Sanjeev Khudanpur$^{1,2}$\thanks{The authors would like to thank Matthew Wiesner for his comments on the paper draft.}}

\address{
  $^1$CLSP \& $^2$HLTCOE, Johns Hopkins University, USA\\
  $^3$NVIDIA Corp., USA\\ 
  $^4$Xiaomi Corp., China}
  
\begin{document}
\ninept
\maketitle
\begin{abstract}
Training automatic speech recognition (ASR) systems requires large amounts of well-curated paired data. However, human annotators usually perform ``non-verbatim'' transcription, which can result in poorly trained models. In this paper, we propose \textit{Omni-temporal Classification} (OTC), a novel training criterion that explicitly incorporates label uncertainties originating from such weak supervision. This allows the model to effectively learn speech-text alignments while accommodating errors present in the training transcripts. OTC extends the conventional CTC objective for imperfect transcripts by leveraging weighted finite state transducers. Through experiments conducted on the LibriSpeech and LibriVox datasets, we demonstrate that training ASR models with OTC avoids performance degradation even with transcripts containing up to $70\%$ errors, a scenario where CTC models fail completely. Our implementation is available at \url{https://github.com/k2-fsa/icefall}.

\end{abstract}

\section{Introduction}
\label{sec:intro}

The development of automatic speech recognition (ASR) systems based on neural networks typically requires acquiring vast amounts of accurately transcribed speech~\cite{Li2020OnTC, Lu2020ExploringTF, Wang2020TransformerIA}. However, with the exception of a few very carefully human-annotated or read speech corpora --- the creation of which is highly time-consuming and labor-intensive --- most datasets contain errors that might degrade the ASR performance of trained models, especially when the amount of data is relatively limited. This poses a major challenge to the development of accurate ASR systems outside large industrial groups.

Existing methods to address this issue may be categorized into two perspectives: \textit{data} and \textit{model}. Data cleaning is a broadly used technique wherein errors in datasets are detected and discarded, usually through multiple stages of model training and alignment. However, data cleaning is usually done at the utterance level~\cite{Driesen2013LightlySA, Long2013ImprovingLS, Lanchantin2016SelectionOM, wiesner2021training, xu-koehn-2017-zipporah}; i.e., utterances in long recordings that do not align perfectly with the annotations are often discarded. Such a process can potentially filter out a large number of imperfect utterances that may still contain a lot of useful information for the model. Table~\ref{table:data_filtering} presents the summary statistics of some popular ASR datasets that were created using such a data-cleaning process from sources containing non-verbatim transcripts. Unsupervised training~\cite{sedom, wav2vecu, Liu2022TowardsEU, Klejch2021DecipheringSA, Gao2022EUROEU}, on the other hand, leverages non-parallel speech and text, where models are designed to learn high-level features from audio. However, training such models is often challenging due to the lack of supervision, and the improvements achieved by unsupervised training tend to be limited.

From a model perspective, \emph{weakly supervised} learning tackles data errors by designing models that can detect and learn from useful portions of the data while avoiding contamination from errors. This class of methods includes the recently proposed star temporal classification (STC)~\cite{stc}, wild-card CTC (W-CTC)~\cite{wctc}, alternative
pseudo-labeling (APL)~\cite{apl}, and bypass temporal classification (BTC)~\cite{gao2023bypass}. These methods enable the use of audio data combined with non-verbatim transcripts, such as videos with closed captions, which are widely available and accessible on the internet. Our proposed OTC training objective falls into this model perspective, and we will describe it in Section~\ref{sec:otc}.

\begin{table}[t]
\caption{Data filtering statistics for popular ASR corpora. Raw and filtered refer to the amount (in hours) of source audio and the final prepared data, respectively.}
\label{table:data_filtering}
\centering
\adjustbox{max width=\linewidth}{
\begin{tabular}{@{}lrrr@{}}
\toprule
\textbf{Dataset}        & \textbf{Raw (h)}          &\textbf{Filtered (h)}  & \textbf{Ratio ($\%$)}\\ 
\midrule
TED-LIUM~\cite{rousseau-etal-2012-ted}              & 216                & 118          & 54.6   \\ 
TED-LIUM 2~\cite{rousseau-etal-2014-enhancing}      & 351                & 207          & 58.9     \\ 
TED-LIUM 3~\cite{hernandez2018ted}                  & 540                & 452          & 83.7     \\ 
Europarl-ST~\cite{9054626}                          & 816                & 255          & 31.3     \\
HK-LegiCoST~\cite{Xiao2023HKLegiCoSTLN}             & 1,400              & 609          & 43.5   \\ 
Europarl-ASR~\cite{diazmunio21_interspeech}         & 4,900              & 1,300        & 26.5       \\ 
\bottomrule
\end{tabular}}
\end{table}

Data and model based methods may also be used in combination. For instance, Wav2Vec-U~\cite{wav2vecu} used self-supervised training on unlabeled speech. Meta's MMS~\cite{Pratap2023ScalingST} used multiple rounds of data alignment using a model trained with the STC objective. The recently released Whisper~\cite{Radford2022RobustSR} leveraged 680k hours of weakly filtered web data to achieve impressive ASR performance.

In this paper, we focus on the model perspective of weakly supervised ASR. We argue that in order to train models on reasonably-sized corpora with imperfect transcripts (e.g. non-verbatim annotations), the model needs to explicitly account for errors  (i.e., substitution, insertion, and deletion errors) that exist within the transcripts during the training process.\footnote{Such label errors may be relatively less hazardous in the very-large-data regime (such as Whisper), but most academic (and industrial) systems trained on medium scale data, often for copyright/privacy reasons. At this scale, erroneous transcripts may be devastating (\S~\ref{sec:librivox}).} Our observation is inspired by numerous recent studies such as STC, W-CTC, APL, and BTC, as mentioned earlier. These existing methods focus on specific categories of transcript errors: STC and W-CTC tackle deletion errors, APL addresses substitution errors, and BTC handles both substitution and insertion errors. However, the specific type of error present in real-world data sources cannot be predetermined and may include all of these. Consequently, these are only partial solutions, and it is imperative to develop a method capable of effectively handling and accommodating all types of errors.

\begin{figure*}[t]
    \begin{subfigure}{\linewidth}
    \centering
    \includegraphics[width=0.65\linewidth]{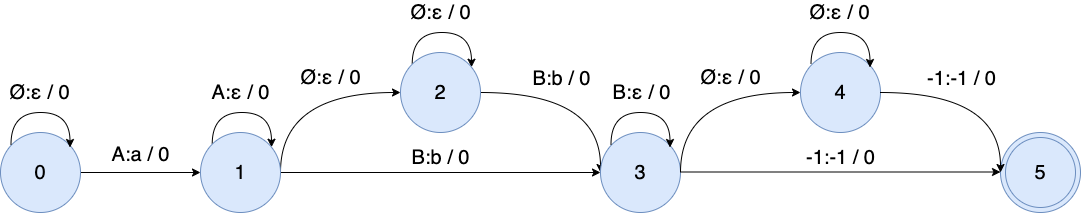}
    \caption{CTC training graph.}
    \label{fig:ctc_training_graph}
    \end{subfigure}\hfill
    \begin{subfigure}{\linewidth}
    \centering
    \includegraphics[width=0.95\linewidth]{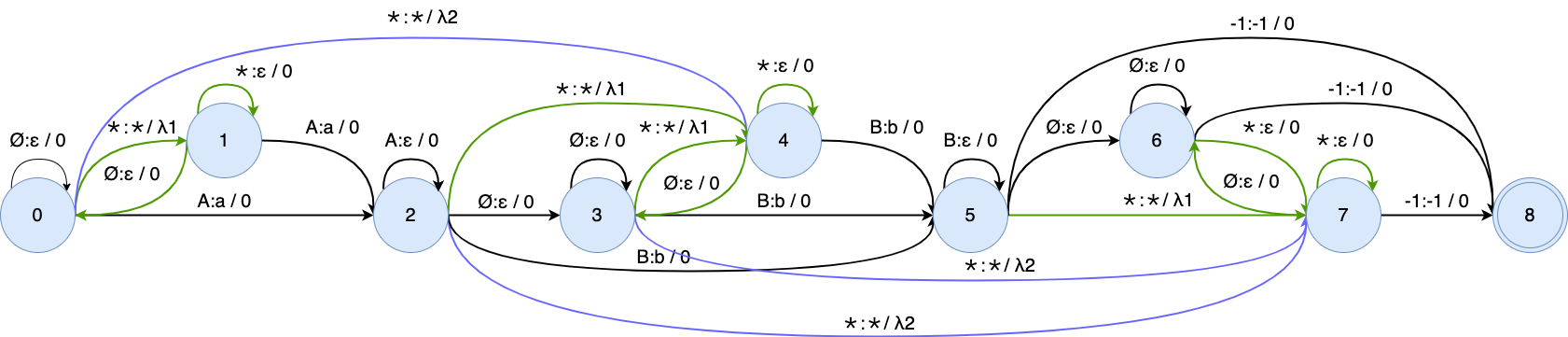}
    \caption{OTC training graph. The self-loop arcs and bypass arcs are highlighted in \textcolor{our_green}{green} and \textcolor{blue1}{blue}, respectively.}
    \label{fig:otc_training_graph}
    \end{subfigure}
    \caption{Training graphs of the transcript ``a b" given the lexicon \{a:A, b:B\}. A state labeled `0' is the starting state. The state with the double circle (state 5) is the final state. States are connected by directed arcs. Each arc has an input and output symbol (separated by a colon) and a weight (after a slash). The arc labeled `-1' is a special arc pointing to the final state.}
    \label{fig:training_graph}
\end{figure*}

To this end, we propose a criterion called Omni-temporal Classification (OTC), an extension of Connectionist Temporal Classification (CTC)~\cite{ctc}.
OTC integrates the advancements made in previous works such as STC and BTC~\cite{stc, gao2023bypass} and supersedes them by addressing all types of errors encountered during training (hence the name ``omni'').
We achieve this by directly encoding label uncertainties, encompassing substitution, insertion, and deletion errors, into the training graph by leveraging the weighted finite-state transducer (WFST) framework.
Our implementation is built on the k2 toolkit\footnote{\url{https://github.com/k2-fsa/k2}}, which provides GPU-based implementations of all WFST operations to enable faster training.
Through extensive experiments on the LibriSpeech and LibriVox corpora, we demonstrate that OTC-based training maintains the performance of the ASR model despite erroneous transcripts. Notably, on the LibriSpeech dataset, when the proportion of synthetic errors exceeds 50\%, OTC-based models still obtain 10-30\% word error rates (WERs), whereas CTC training fails completely.
Additionally, we show that our method is robust against under-processed data of lower quality, which requires minimal human intervention.

\section{Preliminaries}
\label{sec:preliminaries}

\subsection{ASR with CTC}

Given an acoustic feature sequence $\mathbf{x} = [x_{1},\dots,x_{T}]$ of length $T$, where $x_t \in \mathbb{R}^d$ corresponds to the $d$-dimensional, real-valued acoustic representation at frame $t$, an ASR system predicts the most ``probable'' transcript $\hat{\mathbf{y}}$, i.e.,
\begin{equation}
   \hat{\mathbf{y}} = \argmax_{\mathbf{y}} P(\mathbf{y} | \mathbf{x}).
\end{equation}

Here, $\mathbf{y} = [y_{1},\dots, y_{U}]$ has length of $U$, where $y_{u} \in \mathcal{V}$ represents a discrete unit from a finite vocabulary $\mathcal{V}$.
CTC models the posterior probability $P(\mathbf{y} | \mathbf{x})$ by marginalizing over all possible frame-level alignments $\bm{\pi} = [\pi_{1}, \dots, \pi_{T}]$ between $\mathbf{x}$ and $\mathbf{y}$. Each $\pi_{t} \in \mathcal{V} \cup \oslash$, where $\oslash$ is a special symbol used in CTC to align extra acoustics, given that $T \geq U$. CTC further assumes that the output units in the frame-level alignment $\bm{\pi}$ are {\em conditionally independent} given the acoustic features $\mathbf{x}$, which yields

\begin{align}
       P(\mathbf{y} | \mathbf{x}) &= \sum_{\bm{\pi} \in \mathcal{B}^{-1}(\mathbf{y})} P(\mathbf{y}, \bm{\pi} | \mathbf{x}) =  \sum_{\bm{\pi} \in \mathcal{B}^{-1}(\mathbf{y})} P(\bm{\pi} | \mathbf{x}) \\
       &= \sum_{\bm{\pi} \in \mathcal{B}^{-1}(\mathbf{y})} \prod_{t=1}^{T} P(\pi_{t} | \mathbf{x}),
\end{align}
where $\mathcal{B}$ is a deterministic mapping from the alignment sequence $\bm{\pi}$ to transcript $\mathbf{y}$ by removing $\oslash$ and merging adjacent repetitions.

\subsection{CTC within the WFST framework}

A weighted finite-state transducer (WFST) is a directed graph that represents a function that maps sequences of input symbols to output symbols. Each edge (or arc), denoted as $e$, is assigned a weight, $\omega(e)$. Two WFSTs $H_{1}$ and $H_{2}$ can be composed to cascade mapping operations, denoted as $H_{1} \circ H_{2}$. In the resulting WFST, the weight of each edge is the $\otimes$-product of the corresponding edge weights from the source WFSTs. When training ASR models, we use WFSTs on the log semiring, such that edge weights may be interpreted as log probabilities; consequently, $\otimes$ is simply the addition operation.\footnote{Refer to \cite{Mohri2008SpeechRW} for details about WFSTs for speech recognition.}

Previous studies~\cite{stc, gao2023bypass, Miao2015EESENES, Sak2015LearningAF, Laptev2021CTCVT} have shown that the computation of $ P(\mathbf{y} | \mathbf{x})$ can be efficiently implemented within the WFST framework.
$\mathcal{B}^{-1}(\mathbf{y})$, which encompasses all possible alignments for a given $\mathbf{y}$, can be represented by a WFST denoted as $S(\mathbf{y})$. This WFST, often referred to as the \emph{training graph}, is illustrated in Figure~\ref{fig:ctc_training_graph}.
Each path within $S(\mathbf{y})$ corresponds to a distinct alignment $\bm{\pi}$. 
By composing $S(\mathbf{y})$ with the emission WFST $E(\mathbf{x})$ that represents $\log P(\bm{\pi} | \mathbf{x})$ (Fig.~\ref{fig:emission} (a)), $\log P(\mathbf{y} | \mathbf{x})$ can be expressed as

\begin{equation}
    \log P(\mathbf{y} | \mathbf{x}) = 
    \sum_{\bm{\pi} \in S(\mathbf{y})^{-1}}
    \underbrace{P(\bm{\pi} | \mathbf{x})}_{
        \textstyle
            \begin{gathered}
                E(\mathbf{x})
            \end{gathered}
            } \\
    = \text{Weight}(E(\mathbf{x}) \circ S(\mathbf{y})).
\end{equation}

The training graph $S(\mathbf{y})$ can be further factored as  $S(\mathbf{y}) = T \circ L \circ G(\mathbf{y})$, where $T$ removes the special symbol $\oslash$ and merges repeated output units, $L$ maps unit sequences (e.g., BPE or phone) to words, and 
$G(\mathbf{y})$ is a linear WFST representing the transcript $\mathbf{y}$ (Fig.~\ref{fig:ctc_g}). This decomposition yields
\begin{equation}
   \log P(\mathbf{y} | \mathbf{x}) = \text{Weight}(E(\mathbf{x}) \circ T \circ L \circ G(\mathbf{y})).
\end{equation}

In the following sections, we will describe our modifications to $E(\mathbf{x})$ and $G(\mathbf{y})$ that constitute the OTC loss.


\section{Omni-temporal Classification}
\label{sec:otc}

\subsection{WFST topology}

As mentioned earlier, previous CTC variants such as STC~\cite{stc}, W-CTC~\cite{wctc}, and BTC~\cite{gao2023bypass} partially solved the problem of training with imperfect transcripts. The key insight was to extend the training graph by introducing a special token $\star$ that can handle mismatched acoustics. OTC further enhances the flexibility of the training graph to accommodate all potential errors present in the transcript $\mathbf{y}$, including substitution, insertion, and deletion errors. 

\begin{figure}[t]
    \centering
    \hfill
    \begin{subfigure}[b]{0.3\linewidth}
        \centering
        \includegraphics[width=0.9cm]{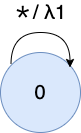}
        \caption{self-loop arc}
        \label{fig:self_loop}
    \end{subfigure}
    \hfill
    \begin{subfigure}[b]{0.4\linewidth}
        \centering
        \includegraphics[width=3.0cm]{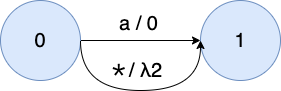}
        \caption{bypass arc}
        \label{fig:bypass}
    \end{subfigure}
    \hfill
    \vspace{2mm}
    \begin{subfigure}[b]{0.9\linewidth}
        \centering
        \includegraphics[width=\linewidth]{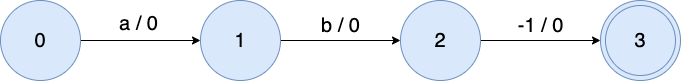}
        \caption{CTC $G(\mathbf{y})$}
        \label{fig:ctc_g}
    \end{subfigure}
    \begin{subfigure}[b]{0.9\linewidth}
        \centering
        \includegraphics[width=\linewidth]{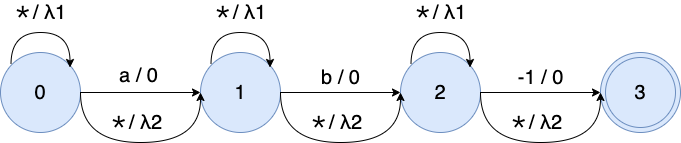}
        \caption{OTC $G_{\text{otc}}(\mathbf{y)}$}
        \label{fig:otc_g}
    \end{subfigure}
\caption{CTC and OTC WFST representations of $\mathbf{y}$: $G(\mathbf{y})$ and $G_{\text{otc}}(\mathbf{y})$. The self-loop arc and bypass arc are associated with penalty $\lambda_{1}$ and $\lambda_{2}$, respectively. An arc with a single symbol indicates identical input and output.}
\end{figure}

\begin{figure}[t]
    \centering
    \includegraphics[width=0.8\linewidth]{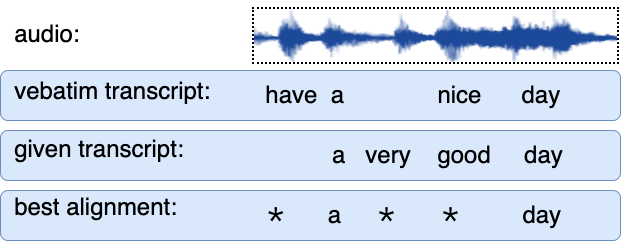}
    \caption{Example demonstrates how OTC tackles errors in the given transcript ``a very good day". The errors include one deletion (missing ``have"), one insertion (``very"), and one substitution (replacing "nice" with "good") error compared to the verbatim transcript. During training, OTC automatically aligns the audio with the pattern " $\star$ a $\star$ $\star$ day", by assigning a higher probability to this alignment.}
    \label{fig:alignment_example}
\end{figure}

To achieve this, we modify $G(\mathbf{y})$ by adding self-loop arcs (Fig.~\ref{fig:self_loop}) into each state and bypass arcs (Fig.~\ref{fig:bypass}) into each arc. Each added arc is associated with the special token $\star$ and penalty $\lambda$ adopted from~\cite{stc, gao2023bypass}. The modified WFST $G_{\text{otc}}(\mathbf{y})$ is shown in Fig.~\ref{fig:otc_g}. This $\star$ symbol acts as a token of uncertainty, offering a preferable choice over aligning with potentially incorrect tokens.
Fig.~\ref{fig:alignment_example} illustrates an example of how the $\star$ token functions.

\begin{figure}[t]
    \centering
    \begin{minipage}[b]{0.9\linewidth}
        \centering
        \centerline{\includegraphics[width=8cm]{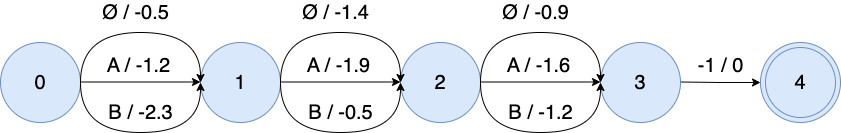}}
        \vspace{2mm}
        \centerline{(a) CTC emission WFST $E(\mathbf{x})$}\medskip
    \end{minipage}
    \begin{minipage}[b]{0.9\linewidth}
        \centering
        \centerline{\includegraphics[width=7.5cm]{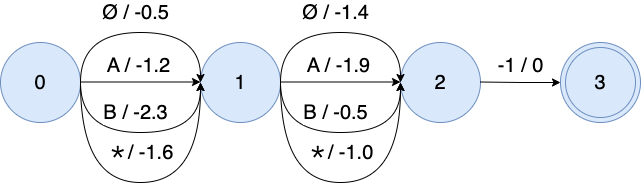}}
        \vspace{2mm}
        \centerline{(b) OTC emission WFST $E(\mathbf{x})$. }\medskip
    \end{minipage}
    \caption{OTC and CTC emission WFST $E(\mathbf{x})$ over 2 frames. For OTC, the weight associated with each arc is its log probability. The weight of $\star$ is the log average probability of `A' and `B': $\log \frac{e^{-1.2} + e^{-2.3}}{2} = -1.6$ and $\log \frac{e^{-1.9} + e^{-0.5}}{2} = -1.0$ for 2 frames.}
    \label{fig:emission}
\end{figure}

\vspace{1mm}
\noindent \textbf{Bypass arcs} were proposed in~\cite{gao2023bypass} to address substitution and insertion errors. 
When encountering an incorrectly substituted word ${y_{u}}'$, the bypass arc provides an alternative path parallel to the erroneous word. This allows the model to associate the acoustics with the $\star$ token instead of the erroneous token, effectively reducing the probability $P({y_{u}}' | \bm{x})$. By doing so, the model avoids learning wrong relationships through back-propagation during training.
This strategy also applies to insertion errors, where the transcript $\mathbf{y}$ contains additional words not present in the intended reference. In such cases, the model allocates a minimal amount of acoustics, usually a single frame, from neighboring tokens to the $\star$ token.

\noindent \textbf{Self-loop arcs} are proposed to handle deletion errors. These errors occur when there are acoustics present in the input data $\mathbf{x}$ which do not have a corresponding token $y_{u}$ in the transcript $\mathbf{y}$. To handle deletions, the self-loop arc can insert an arbitrary number of $\star$ tokens. These $\star$ tokens can then be aligned with the ``orphan'' acoustics in the input $x$, filling in the gaps left by the deleted tokens.

After composing the modified WFST $G_{\text{otc}}(\mathbf{y})$ with $L$ and $T$, the OTC training graph is shown in Fig.~\ref{fig:otc_training_graph}.
We incorporate the penalty strategy introduced in~\cite{gao2023bypass} and apply different configurations for the self-loop arc and bypass arc. The penalties are set as
\begin{equation}
    \lambda_{1_{i}} = \beta_{1} * \tau_{1}^{i},\quad \lambda_{2_{i}} = \beta_{2} * \tau_{2}^{i}
\end{equation}
for the $i$-th training epoch. $\beta$ is the initial penalty that encourages the model to rely more on the given transcript at the start of training. 
It decays exponentially by a factor of $\tau \in (0, 1)$, gradually encouraging the model to align speech with $\star$ when getting confused. 

\vspace{-2mm}

\subsection{Modeling $\star$ token}
\label{ssec:modeling_star}
In~\cite{gao2023bypass}, the symbol $\star$ represented an individual token with its own distribution. 
This approach works for substitution and deletion errors, as the bypass arc offers a non-$\star$ token as the reference.  
In effect, it guides the model to treat $\star$ as a ``garbage'' token capable of matching any non-blank tokens by distinguishing between $\star$ and reference token given the acoustic context.

However, in our preliminary experiments, we found that such modeling of $\star$ is not effective for insertion (self-loop) case, as the model may confound the roles of $\star$ and $\oslash$ symbols as there is no token to be compared within the self-loop arc. 
To remedy this, we adopt the strategy used in STC~\cite{stc}, where $\star$ is represented as the average probability of all non-blank tokens. In our WFST-based implementation, this is done by modifying the emission WFST $E(\mathbf{x})$, as shown in Fig.~\ref{fig:emission} (b).

\section{Data preparation}
\label{sec:librivox}

We use the publicly available LibriSpeech and LibriVox corpora for demonstrating OTC. 

\textbf{LibriSpeech}~\cite{librispeech} (LS) train set consists of 960 hours of read speech divided into three subsets: 100 + 360 (clean) and 500 (other). Additionally, it includes two development subsets and two test subsets, each containing 5 hours of speech. 

\begin{figure}[t!]
    \centering
    \includegraphics[width=0.7\linewidth]{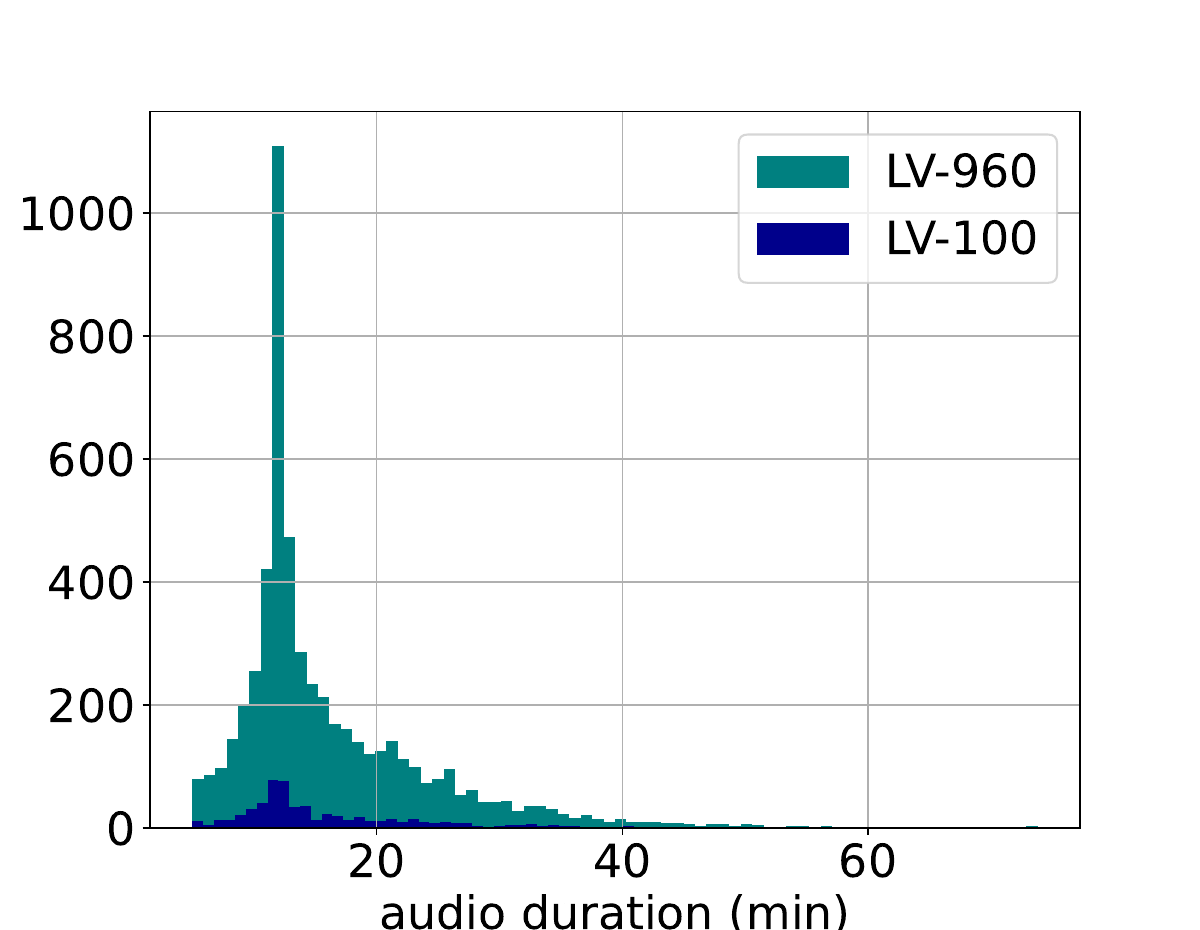}
    \caption{Distribution of chapter audio length of LibriVox 100 hours (\textcolor{lv_green}{green}) and LibriVox 960 hours (\textcolor{lv_blue}{blue}).}
    \label{fig:dur}
\end{figure}

\textbf{LibriVox}\footnote{\url{https://librivox.org}} (LV) is a set of audiobooks, from which the LS dataset is derived. Each audiobook consists of multiple chapters, and the audio is available at the chapter level without any segmentation.
For our investigation, we retrieved the exact portions of the audiobooks that were used to create LS. 
We refer to the retrieved data as \textbf{LV-100} and \textbf{LV-960}, respectively, which correspond to the 100h subset and the full LS, respectively. 
The duration of each chapter varies between 5 minutes and 70 minutes as shown in Fig.~\ref{fig:dur}. It should be noted that the accompanying text derived from the books is not perfect, as it contains errors from human reading.\footnote{Common errors include missing chapter titles and reader information in the text, repetitions in the audio, and reading incorrect words.} 

\subsection{LibriSpeech processing}
\label{ssec:librispeech}
As described in \cite{librispeech}, significant human efforts were dedicated to creating the LS dataset from the LV source.

\begin{enumerate}
\item \textbf{Preprocessing}: The audio of each chapter was divided into segments, with a maximum duration of 30 minutes. These segments were decoded using an external ASR model trained on the VoxForge dataset, using the Kaldi toolkit~\cite{Povey2011TheKS}.

\item \textbf{First alignment (coarse filtering)}: The Smith-Waterman algorithm~\cite{smith1981identification} was applied to compare the decoded text with the original chapter text. This process aimed to identify the region that exhibited the closest text match to retrieve the corresponding speech fragment. The retrieved fragments were subsequently segmented into smaller chunks, ensuring that each chunk was shorter than 35 seconds and separated by periods of silence.

\item \textbf{Second alignment (fine filtering)}: A specialized decoding graph was constructed as a refined filter to only allow arbitrary phone insertions between words in the transcript, or replacement of
words in the transcript. This graph was applied to the smaller audio chunks obtained from the previous stage, in order to filter out chunks whose decoded text significantly differed from the expected transcript, thereby providing a more accurate selection of suitable speech fragments.
\end{enumerate}

\begin{figure}[t!]
    \centering
    \includegraphics[width=1.0\linewidth]{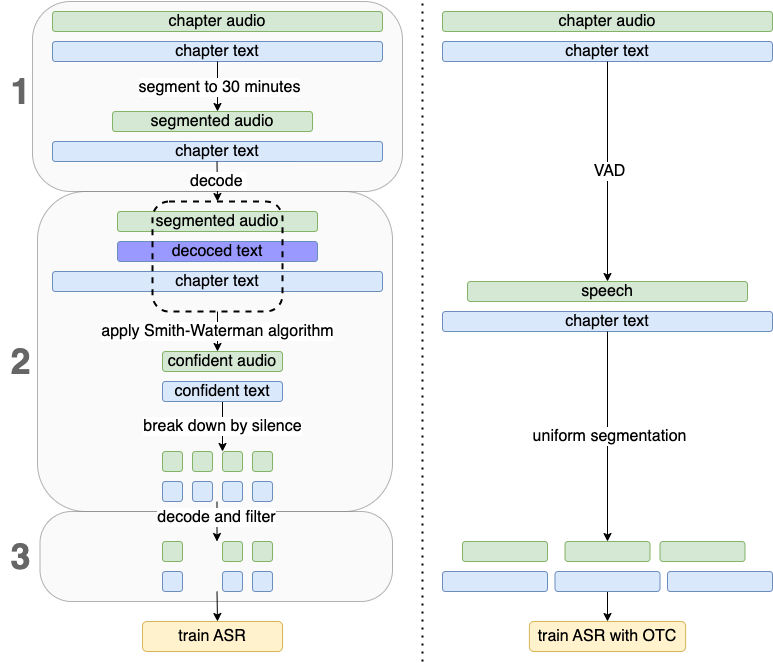}
    \caption{Comparison between data processing of LibriSpeech (left) and LibriVox (right). Audio, reference text, and decoded text are represented as blocks of color: \textcolor{compare_green}{green}, \textcolor{compare_light_blue}{light blue}, and \textcolor{compare_dark_blue}{dark blue}. The 3 big grey blocks on the left represent the following stages: preprocessing, first alignment, and second alignment.}
    \label{fig:librivox}
\end{figure}

\subsection{LibriVox processing: weak supervision}

To demonstrate the efficiency of training an ASR system using OTC on weakly supervised speech data, we use the LibriVox dataset. Our goal is to highlight the minimal effort required for preparing training data in this approach. For a visual representation of the comparison between data preparation for LS and LV, please refer to Fig.~\ref{fig:librivox}.

As described in Section~\ref{sec:librivox}, we extract the chapter-level audio and text from the original audiobooks corresponding to the 100h and 960h sets from LS, and these are referred to as LV-100 and LV-960, respectively. By employing OTC, the data processing workflow can be simplified into two steps. 

\begin{enumerate}
\item First, a voice activity detection (VAD) technique is applied to remove non-speech audio. 
\item Following that, both the audio and text data are uniformly segmented, with each speech segment restricted to a duration of 60 seconds. In Section~\ref{ssec:segmentation_length}, we present an analysis of different choices for this uniform duration.
\end{enumerate}

The processed LV corpus contains errors from two distinct sources. 
The first one, which we label as a ``shift error,'' is a product of uniform segmentation, which can cause a misalignment between the speech and the corresponding text segments. Such a mismatch may occur if the segmented portions do not exactly match the natural boundaries of speech. 
The second error source, which we refer to as an ``internal error,'' originates from the potential imperfections within the original chapter text itself. Such errors are inherent due to human error in the reading process.

\begin{figure}[t]
    \centering
    \includegraphics[width=0.8\linewidth]{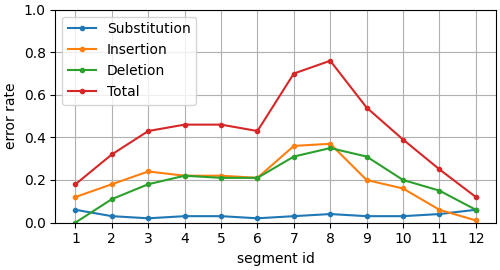}
    \caption{Text error rates of 12 segments within the chapter (id 103-1240), span from 0.2 to 0.8. The text error rates are dominated by shift errors (insertion and deletion), particularly for the segments located in the middle of the chapter.}
    \label{fig:wer}
\end{figure}

In order to gain a clearer understanding of the text mismatch, we manually aligned the text for each segment within a specific chapter (id 103-1240), which we then used as the ground truth.
We compared this ground truth text to the text acquired via uniform segmentation and calculated the word error rates. This was done to numerically express the extent of misalignment, as shown in Fig.~\ref{fig:wer}.

\section{Experimental setup}
\label{sec:experiments}

\subsection{Synthetic data generation}
\label{ssec:synthetic_data}
For a controlled experiment, we first generated synthetic data based on LS \texttt{train-clean-100}. 
This allows us to test whether OTC performs as well as tuning hyper-parameters. 
We introduced substitution, insertion, and deletion, errors by randomly replacing, inserting, and removing each token in the transcript with probabilities $p_{\text{sub}}, p_{\text{ins}}, p_{\text{del}}$, respectively.
We trained models on transcripts containing different degrees of each of these errors, by setting the corresponding probabilities to one of $\{0.1, 0.3, 0.5, 0.7\}$. 
Additionally, we train on transcripts containing all the errors combined. For this, we consider settings where $p_{\text{sub}} + p_{\text{ins}} + p_{\text{del}}$ falls within the same set of values, such that $p_{\text{sub}} = p_{\text{ins}} = p_{\text{del}}$.

\subsection{Model}

We used the wav2vec 2.0 (base) model~\cite{wav2vec2} to extract 768-dimensional features with a stride of 20 ms, from audio recordings originally sampled at 16 kHz. 
For this, we used the extractor available in the S3PRL toolkit~\cite{s3prl}.\footnote{\url{https://github.com/s3prl/s3prl}} These features were fed to an acoustic model based on a 12-layer conformer~\cite{conformer} network. Each conformer block consists of two feed-forward layers, with half-step residual connections. These layers encapsulate multi-headed self-attention and convolution modules, followed by layer normalization. The decoder is simply a linear layer with a softmax, which converts encoder representations into a probability distribution over the extended vocabulary. For CTC-based training, the output units comprise BPE units and a blank token. For OTC, we additionally include the $\star$ token in the output, as described earlier.

\begin{figure}[t]
\centering
    \begin{subfigure}[b]{0.49\linewidth}
        \centering
        \includegraphics[width=\linewidth]{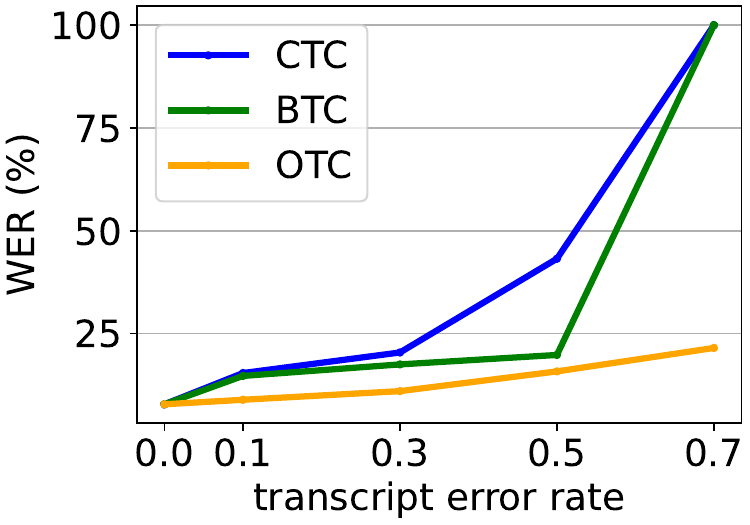}
        \caption{Substitution}
    \end{subfigure}
    \begin{subfigure}[b]{0.49\linewidth}
        \centering
        \includegraphics[width=\linewidth]{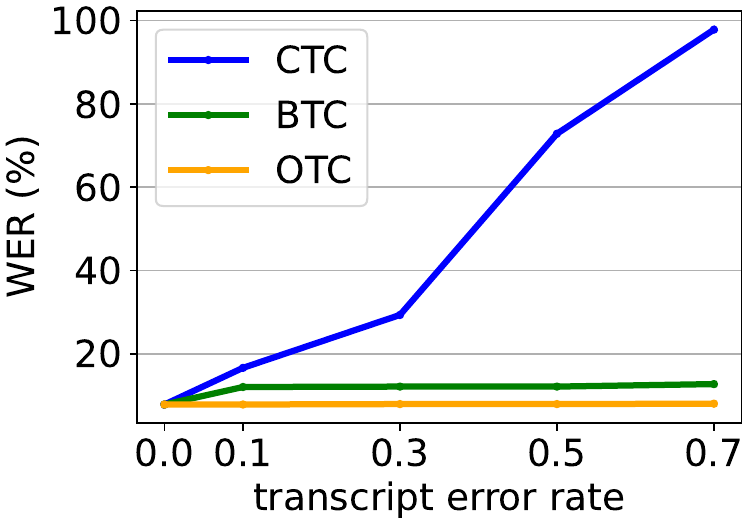}
        \caption{Insertion}
    \end{subfigure}\hfill
    \begin{subfigure}[b]{0.49\linewidth}
        \centering
        \includegraphics[width=\linewidth]{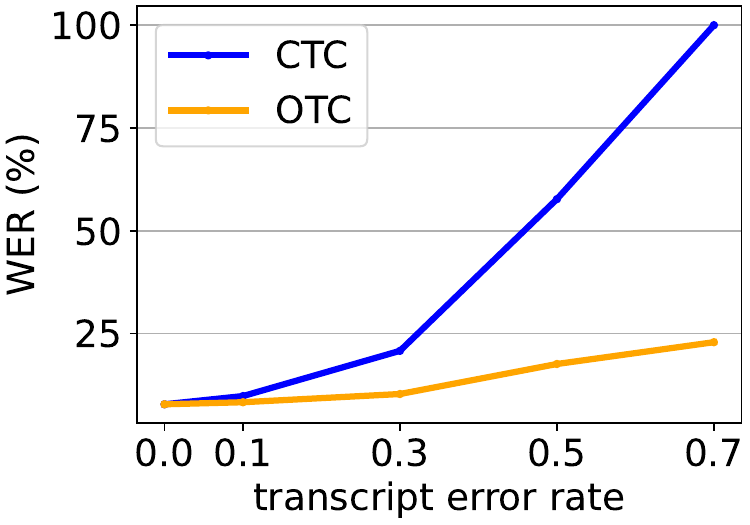}
        \caption{Deletion }
    \end{subfigure}
    \begin{subfigure}[b]{0.49\linewidth}
        \centering
        \includegraphics[width=\linewidth]{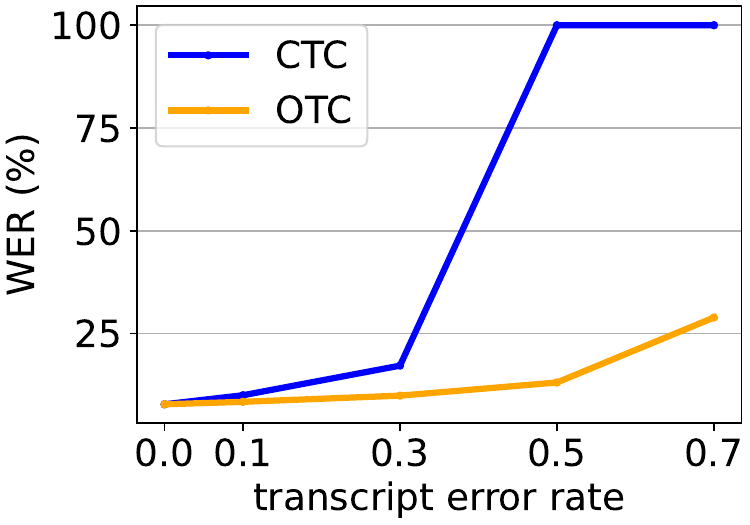}
        \caption{Mixture}
    \end{subfigure}
    \caption{WERs of models trained on LS \texttt{train-clean-100} with synthetic transcript errors: substitution, insertion, deletion, and a mixture of these at error rates: $\{0.0, 0.1, 0.3, 0.5, 0.7\}$. BTC results are compared in (a) substitution and (b) insertion. Results of CTC, BTC, and OTC are depicted in \textcolor{blue}{blue}, \textcolor{olive}{green}, and \textcolor{orange}{orange}, respectively.}
    \label{fig:wer_clean_100}
\end{figure}

\subsection{Tokenization}
We used byte pair encoding (BPE)~\cite{Sennrich2015NeuralMT} for sub-word tokenization. For this, we used the implementation available in the \texttt{sentencepiece} toolkit\footnote{\url{https://github.com/google/sentencepiece}}. We explored vocabulary sizes of 100, 200, and 500, and determined that a size of 200 yields the optimal performance. Further analysis and details can be found in Section~\ref{ssec:vocab_size}.

\section{Results \& Discussion}
\label{sec:results}

\subsection{LibriSpeech 100h}

As mentioned in Section~\ref{ssec:synthetic_data}, we first experimented with OTC-based training for LS \texttt{train-clean-100} by incorporating synthetic errors. 
We performed experiments under four scenarios: substitution, insertion, deletion, and a mix of all three errors at different error rates. 
For each of these weakly supervised settings, we trained models using both CTC and OTC objectives and evaluated their word error rate (WER) performance on LS \texttt{test-clean} using greedy decoding. 
For substitution and insertion cases, we also compared the results with BTC. The comparisons are shown in Fig.~\ref{fig:wer_clean_100}.

As the error rates increase (especially beyond 0.5), the performance of the ASR system using regular CTC worsens, and it fails to converge entirely beyond error rates of 0.7. 
On the other hand, models trained with OTC maintain their performance with only small degradation.
OTC also outperforms BTC in scenarios involving substitution and insertion errors, due to the strategy of modeling the $\star$ token described in Section~\ref{ssec:modeling_star}. Importantly, OTC performs \textbf{on par} with CTC in the case of verbatim transcripts, as shown in Table~\ref{table:no_error}.

\subsection{LibriVox}

We also evaluated OTC on LV-100 and LV-960 datasets. 
To establish a baseline for comparison, we trained the ASR system using the CTC on the LibriVox dataset.
We define a ``topline'' as the result obtained by training the ASR system on the corresponding LS subset, which is clean and well-segmented as outlined in Section~\ref{ssec:librispeech}. 
We decoded using a 3-gram language model, and the results are shown in Table~\ref{table:librivox_librispeech}.
The results obtained using CTC are emphasized in \textcolor{red}{red}. 

These experiments highlight the ineffectiveness of direct training on the LV dataset using the CTC approach due to transcript errors. However, when utilizing OTC for training, we observed a significant reduction in the performance gap between the LS and LV datasets.
Specifically, when considering the 100-hour subset, OTC only incurs a loss of 2.7\% absolute WER on \texttt{test-clean} (from 5.2\% to 7.9\%) and 4.6\% absolute WER \texttt{test-other} (from 13.5\% to 18.9\%). Similarly, on the 960-hour set, OTC experiences a loss of 2.8\% on \texttt{test-clean} (from 3.3\% to 6.1\%) and 4.3\% on \texttt{test-other} (from 8.2\% to 12.5\%).

\begin{table}[t]
\caption{WER ($\%$) of CTC/OTC on LS \texttt{train-clean-100} with verbatim transcript (transcript error rate = 0.0).}
\vspace{2mm}
\label{table:no_error}
\centering
\begin{tabular}{llllll}
\bottomrule
Transcript Error Rate & CTC & OTC & \\ \bottomrule
0.0                    & 7.8                         & 7.8  \\ \bottomrule
\end{tabular}
\end{table}

\begin{table}[t]
\caption{Comparison of performance (WER ($\%$)) of ASR systems trained on LS and LV using CTC/OTC.}
\label{table:librivox_librispeech}
\small
\begin{tabular}{llllll}
\bottomrule
                                              & \multicolumn{1}{c}{}                                & \multicolumn{2}{c}{Dev} & \multicolumn{2}{c}{Test} \\ \cmidrule(r{1pt}){3-4} \cmidrule(l{1pt}){5-6} 
\multirow{-2}{*}{}                   & \multicolumn{1}{c}{\multirow{-2}{*}{Training Data}} & Clean      & Other      & Clean       & Other      \\ \bottomrule
\rowcolor[HTML]{FFCCC9} 
\cellcolor[HTML]{FFCCC9}                      & LS \texttt{train-clean-100}                               & 4.9        & 13.6       & 5.2         & 13.5        \\
\rowcolor[HTML]{FFCCC9} 
\multirow{-2}{*}{\cellcolor[HTML]{FFCCC9}CTC} & LV-100                                        & 98.9       & 98.8       & 98.8        & 98.7        \\
OTC                                           & LV-100                                        & 7.4        & 18.4       & 7.9         & 18.9        \\ \bottomrule
\rowcolor[HTML]{FFCCC9} 
\cellcolor[HTML]{FFCCC9}                      & LS \texttt{train-960}                                     & 3.1        & 8.1        & 3.3         & 8.2          \\
\rowcolor[HTML]{FFCCC9} 
\multirow{-2}{*}{\cellcolor[HTML]{FFCCC9}CTC} & LV-960                                        & 99.2       & 99.8       & 99.6        & 99.7       \\
OTC                                           & LV-960                                        & 5.9        & 12.6       & 6.1         & 12.5       \\ \bottomrule
\end{tabular}
\end{table}

\section{Analysis}
\label{sec:analysis}
The analyses are carried out using the LS \texttt{train-clean-100} with synthetic errors in the transcripts. Here, ``error type'' refers to the specific error made in the transcripts, while ``error rate'' indicates the ratio of that error to the reference transcripts.

\subsection{Impact of penalty}
\begin{figure}[t]
\begin{subfigure}{0.49\linewidth}
\centering
\includegraphics[width=\linewidth]{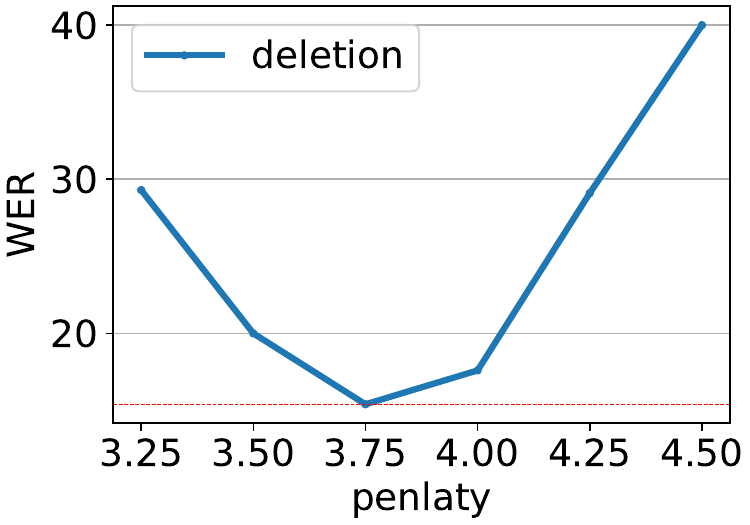}
\caption{Deletion}
\label{fig:del_penalty}
\end{subfigure}
\begin{subfigure}{0.49\linewidth}
\centering
\includegraphics[width=\linewidth]{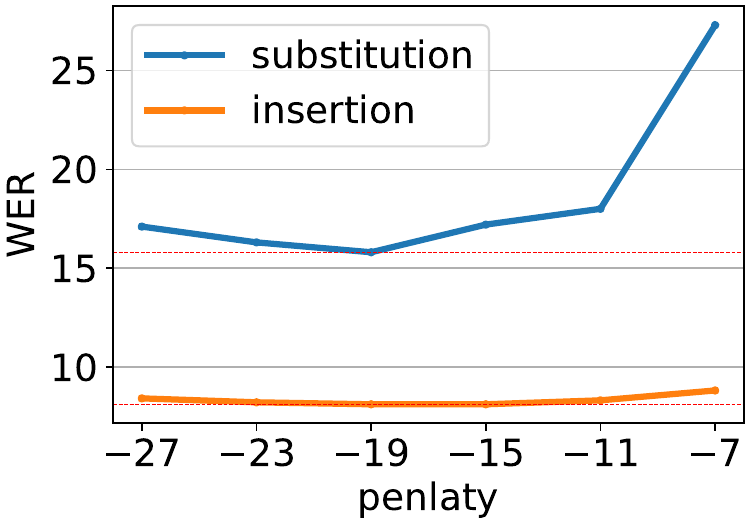}
\caption{Substitution}
\label{fig:sub_penalty}
\end{subfigure}
\vspace{-1em}
\caption{Self-loop and bypass penalty impact on WER ($\%$).}
\label{fig:penalty}
\end{figure}

The initial penalty $\beta_{1}$ and $\beta_{2}$ on the self-loop arc and bypass arc are determined empirically. 
We experimented with a range of different values and evaluated their impact on the performance of the ASR system.
Parameter $\beta_{1}$ was selected for cases involving deletion errors only, where the error rate was set to $0.5$ and the system was configured to only enable self-loop. 
$\beta_{2}$ was chosen for cases of substitution and insertion errors separately, with an error rate again set to $0.5$, enabling only the bypass arc.
The decay factors are set to $\tau_{1} = 0.999, \tau_{2} = 0.975$. 
The comparison is shown in Fig.~\ref{fig:penalty}

\begin{table}[t]
\caption{WER ($\%$) for substitution/deletion at rate 0.5 across subword vocabulary size. The best result is highlighted in \textbf{bold}.}
\vspace{2mm}
\label{table:vocab_size}
\centering
\begin{tabular}{llllll}
\bottomrule
\multirow{2}{*}{Error Type} & \multirow{2}{*}{Error Rate} & \multicolumn{3}{c}{Vocabulary Size} &  \\ \cline{3-5}
                            &                             & 100     & 200              & 500    &  \\ \bottomrule
substitution                & 0.5                         & 15.8    & \textbf{15.2}    & 16.4   &  \\
deletion                    & 0.5                         & 16.6    & \textbf{15.4}    & 18.9   &  \\ \bottomrule
\end{tabular}
\end{table}


In the deletion case, when $\lambda_{1}$ is set to 3.75, the system achieves its lowest WER of 15.4.
In the substitution and insertion case, the system achieves its best performance with a WER of 15.8 for substitution and 8.1 for insertion when $\beta_{2}$ is set to -19.

\subsection{Segmentation length choices}
\label{ssec:segmentation_length}
Traditional ASR training typically utilizes utterances with lengths under 30 seconds, achieving optimal performance within the 10 to 15-second range~\cite{Valenta2014OnTI}. However, in the context of OTC, uniform segmentation with short lengths introduces significant ``shift errors'' that misalign speech and text, while excessively long segments can overwhelm GPU memory in the conformer architecture. We conducted an exploration of different segment durations, including 15, 30, and 60 seconds. Our findings revealed that the model fails to converge when using 15 or 30-second segments. However, we observed that utilizing 60-second segments strikes the optimal balance.

\subsection{Subword vocabulary size}
\label{ssec:vocab_size}
We conducted experiments using subword vocabularies of sizes 100, 200, and 500, focusing on cases of substitution and deletion errors with an error rate of 0.5. The results, shown in Table~\ref{table:vocab_size}, demonstrate that a vocabulary size of 200 consistently yielded the best performance in terms of WER for both substitution and deletion errors (namely, 15.2\% for substitution and 15.4\% for deletion).

%

\section{Conclusion}
\label{sec:conclusion}

We proposed a novel training criterion, Omni-Temporal Classification (OTC), for training ASR systems using weakly supervised data, such as non-verbatim transcripts. OTC allows the model to learn speech-text alignment while effectively addressing errors present in the transcripts within the WFST framework. Experimental results on the LibriSpeech and LibriVox datasets demonstrated that models trained with OTC maintain reasonable ASR performance even when the transcripts contain up to $70\%$ errors of different types. From the LibriVox experiments, it is evident that our method can significantly reduce the human effort required for data preparation in training ASR systems.

\section{Acknowledgement}
\label{sec:acknowledgement}

This work was partially supported by the National Science Foundation CCRI program via Grant No 2120435.

\clearpage

\bibliographystyle{IEEEbib}
\small
\bibliography{refs}

\end{document}